\documentclass[12pt,preprint]{aastex}

\shorttitle{The Lorentz factor distribution and luminosity
function of relativistic jets in AGNs} \shortauthors{Liu \& Zhang}

\begin{document}

%% LaTeX will automatically break titles if they run longer than
%% one line. However, you may use \\ to force a line break if
%% you desire.

\title{The Lorentz factor distribution and luminosity function of relativistic jets in AGNs}

%% Use \author, \affil, and the \and command to format
%% author and affiliation information.
%% Note that \email has replaced the old \authoremail command
%% from AASTeX v4.0. You can use \email to mark an email address
%% anywhere in the paper, not just in the front matter.
%% As in the title, use \\ to force line breaks.

\author{Yuan Liu\altaffilmark{1}, Shuang Nan Zhang\altaffilmark{1,} \altaffilmark{2,} \altaffilmark{3}}

%% Notice that each of these authors has alternate affiliations, which
%% are identified by the \altaffilmark after each name.  Specify alternate
%% affiliation information with \altaffiltext, with one command per each
%% affiliation.

\altaffiltext{1}{Physics Department and Center for Astrophysics,
Tsinghua University, Beijing, 100084, China
(zhangsn@tsinghua.edu.cn, yuan-liu@mails.tsinghua.edu.cn)}
\altaffiltext{2}{ Key Laboratory of Particle Astrophysics,
Institute of High Energy Physics, Chinese Academy of Sciences,
Beijing, China} \altaffiltext{3}{ Physics Department, University
of Alabama in Huntsville, Huntsville, AL 35899, USA
(zhangsn@uah.edu)}

%% Mark off your abstract in the ``abstract'' environment. In the manuscript
%% style, abstract will output a Received/Accepted line after the
%% title and affiliation information. No date will appear since the author
%% does not have this information. The dates will be filled in by the
%% editorial office after submission.

\begin{abstract}
The observed apparent velocities and luminosities of the relativistic jets in AGNs are
significantly different from their intrinsic values due to strong special relativistic
effects. We adopt the maximum likelihood method to determine simultaneously the
intrinsic luminosity function and the Lorentz factor distribution of a sample of AGNs.
The values of the best estimated parameters are consistent with the previous results,
but with much better accuracy. In previous study, it was assumed that the shape of the
observed luminosity function of Fanaroff-Riley type II radio galaxies is the same with
the intrinsic luminosity function of radio loud quasars. Our results prove the validity
of this assumption. We also find that low and high redshift groups divided by $z=0.1$
are likely to be from different parent populations.
\end{abstract}

%% Keywords should appear after the \end{abstract} command. The uncommented
%% example has been keyed in ApJ style. See the instructions to authors
%% for the journal to which you are submitting your paper to determine
%% what keyword punctuation is appropriate.


%% From the front matter, we move on to the body of the paper.
%% In the first two sections, notice the use of the natbib \citep
%% and \citet commands to identify citations.  The citations are
%% tied to the reference list via symbolic KEYs. The KEY corresponds
%% to the KEY in the \bibitem in the reference list below. We have
%% chosen the first three characters of the first author's name plus
%% the last two numeral of the year of publication as our KEY for
%% each reference.

%% Authors who wish to have the most important objects in their paper
%% linked in the electronic edition to a data center may do so by tagging
%% their objects with \objectname{} or \object{}.  Each macro takes the
%% object name as its required argument. The optional, square-bracket
%% argument should be used in cases where the data center identification
%% differs from what is to be printed in the paper.  The text appearing
%% in curly braces is what will appear in print in the published paper.
%% If the object name is recognized by the data centers, it will be linked
%% in the electronic edition to the object data available at the data centers
%%
%% Note that for sources with brackets in their names, e.g. [WEG2004] 14h-090,
%% the brackets must be escaped with backslashes when used in the first
%% square-bracket argument, for instance, \object[\[WEG2004\] 14h-090]{90}).
%%  Otherwise, LaTeX will issue an error.

\section{Introduction}
%\citep{hen61,lyn68,spi85},  \citet{aur82}

The large scale and relativistic jets are important
characteristics of active galactic nuclei (AGNs). Determining
their proper velocity distribution and intrinsic luminosity
function is of fundamental importance for understanding the
central engine of AGNs. However, the observed apparent velocities
and luminosities of the relativistic jets in AGNs are
significantly different from their intrinsic values due to strong
special relativistic effects, i.e. $\beta _{app}  = \frac{{\beta
\sin \theta }}{{1 - \beta \cos \theta }}$ and $S = \frac{{L\delta
^p }}{{4\pi d_L^2 (z)}}(1 + z)^{1 + \alpha } $, where $\beta
,\;\beta _{app} ,\;\theta ,\;L,\;S,\;\delta \;{\rm{and}}\;d_L $
 are the proper velocity, the apparent
velocity, the angle between the axis of jet and the line of sight, the intrinsic
luminosity, the observed flux, the Doppler factor and the luminosity distance,
respectively. The index $p$ is equal to $n-\alpha$ , where $\alpha$ is the spectral
index ($S \propto \nu ^\alpha  $) and  $n=2, 3$ for continuous or discrete jet,
respectively.

Due to the lack of reliable and accurate methods to obtain $\beta$
or $\theta$ of each source, it remains difficult to determine
directly the intrinsic luminosity and the proper velocity of a
relativistic jet. Several authors have addressed these problems
with statistical methods. For example, Padovani \& Urry (1992)
determined the evolution parameters of different types of galaxies
with the $<V/V_m>$ method, and then used the observed luminosity
function of Fanaroff-Riley type II radio galaxies (FR II) as the
intrinsic luminosity function of relativistic jets in AGNs, after
multiplying a constant factor. This approach is at best an
approximation only, because the observed luminosity function of FR
II may or may not resemble the intrinsic luminosity function of
relativistic jets in AGNs. Lister \& Marscher (1997) extended this
work with a simulation approach by fully considering the flux
limit and Doppler beaming effect. However, they utilized the
parameters determined by Padovani \& Urry (1992) due to the size
limit of the simulation. Therefore, the above mentioned approaches
could not determine simultaneously the intrinsic luminosity and
Lorentz factor distribution of relativistic jets in AGNs.
Actually, the observed luminosity function and the apparent
velocity distribution can be reproduced if the intrinsic
luminosity and the Lorentz factor distribution are given (Lister
2003, Vermeulen \& Cohen 1994). Thus the intrinsic luminosity and
Lorentz factor distribution could be inferred simultaneously with
the maximum likelihood method. Therefore, we could test the
previous assumption that the observed luminosity function of FR II
has the same shape with the intrinsic luminosity function of the
radio loud quasars.

The form of the maximum likelihood estimator could be written as
 \[
MLE = \sum\limits_i {C\ln p(\beta _{app,i} \;,\;L_{o,i} \;,\;z_i
)} ,\] where $i$ denotes the number of each source in the sample
and $p$ is the probability density of detecting a source for the
given apparent velocity, observed luminosity and redshift. $C$ is
the normalized factor. The probability density $p$ is
 \[
p(\beta _{app} ,L_o ,z) = f(\beta _{app} |\;L_o ,\;z)\Phi (L_o
,\;z)d_A (z)^2 (1 + z)^3 c\frac{{dT(z)}}{{dz}} ,\]
 where $\beta_{app}$, $d_A$, $T(z)$, $ f(\beta _{app} |\;L_o ,\;z)$ and $\Phi (L_o ,z)$
 are the apparent velocity,
the angular distance, the look-back time, the apparent velocity distribution (for the
given observed luminosity and redshift) and the observed luminosity function,
respectively. As proved by Vermeulen \& Cohen (1994, also see Appendix A at the end of
this paper), if we assume the intrinsic luminosity function is a single power law
($\Phi (L) \propto L^{ - A}$), $L_o$ will not appear in $ f(\beta _{app} |\;L_o ,\;z)$.
Therefore, we could deal with the apparent velocity distribution and the observed
luminosity function separately.

We present the results obtained from the apparent velocity and the
observed luminosity (also the redshift) in \S2. The implication of
our results for the unification scheme of radio loud AGNs is
addressed in \S3. In \S4 we summarize our conclusions and make
some discussions. The cosmological parameters adopted throughout
this paper are $(H_0 ,\;\Omega _M ,\;\Omega _\Lambda  ) =
(75\;{\rm{km/s/Mpc}},\;0.27,\;0.73)$ .

\section{Sample selection and data analysis}\label{section2}
To investigate the real parent population of the radio loud AGNs,
we should adopt a large sample containing different kinds of radio
loud AGNs, including radio galaxies, quasars, and blazars. The
largest available sample of the kinetics of relativistic jets in
AGNs is from the ``VLBA 2 cm survey" (Kellermann et al. 2004). Its
successor, the ``MOJAVE survey
\footnote[1]{http://www.physics.purdue.edu/astro/MOJAVE}",
provides a complete flux-limited sample further. Due to the high
frequency observation (15 GHz), any radio emission from large
scale structures is effectively excluded. The majority of the
sample are the flat-spectrum radio quasars (FSRQ), which is the
most luminous type in AGNs. According to the unified model, FSRQ
is the beamed version of FRII due to the small angle between the
jet and the line of sight.

\subsection{Analysis of the apparent velocity data}

In this section we only make use of the apparent velocity data to determine the
intrinsic luminosity and Lorentz factor distribution. The detailed formulae of the
apparent velocity distribution are presented in Appendix A.

Our sample of the apparent velocity data comes from the ``VLBA 2 cm survey" (Kellermann
et al. 2004). We only adopt the fastest component of each source that has a quality
factor of ``Good" or ``Excellent" (see Figure 11 in Kellermann et al. 2004), since
several authors have pointed out that the velocities of some patterns are much slower
than the bulk velocity responsible for the Doppler boosting effect (Cohen et al. 2006).
The sample used here contains 16 BL Lac objects, 12 radio galaxies and 76 quasars. We
assume the Lorentz factor distribution is also a single power law ($p(\gamma ) \propto
\gamma ^a \;(1<\gamma<\gamma_{\max })$) as adopted by previous studies. As indicated by
previous studies, the maximum value of $\beta_{app}$ is nearly the same as
$\gamma_{max}$ (Vermeulen \& Cohen 1994). The maximum value of
 $\beta_{app}$ in the sample is about 31, and therefore we set $\gamma _{\max }=32$
 for safety. As
assumed by the previous study (Lister \& Marscher 1997), we set $n=2$ and $\alpha=0$.
It has been found that these simple assumptions could give reasonable results. The
estimated result from the whole sample is shown in Figure 1. Obviously, the constraint
on the parameters is weak, which is due to the limitation of the observed data.
However, it is still useful to constrain the values of parameters with such data,
because this allows us to compare between the results from the apparent velocity data
and that obtained from the observed luminosity data, in order to check the consistency
of our models.

By fixing the index $A$, we could investigate whether the Lorentz
factor
 distribution evolves with the redshift. We fix  $A=3.10$ (the best estimated value of
 $A$ for the whole sample) and estimate $a$ in each redshift bin. As shown in Figure 2, the Lorentz
 factor distribution is much steeper in the first bin ($0<z<0.1$) than in other bins.
 The difference of the apparent velocity distributions between the first bin
 and other bins is also significant at the 99.98\% level obtained by K-S test.
 Therefore, we should divide the sample into low and high redshift groups by $z=0.1$.
 The estimated parameters of the two groups are also shown in Figure 1.
 The difference between the results from the whole sample and the high redshift
 sample is only slight, whereas the confidence region of low redshift sources differs with others at 90\% confidence.

\subsection{Analysis of the observed luminosity function and
redshift distribution}

We assume the local intrinsic luminosity function is a single
power law
 \[\Phi (L,\;z = 0) = kL^{ - A} \;\;(L_1  \le L \le L_2 ),\]
and to compare with the previous results, we also assume a pure luminosity evolution
model as used in Padovani \& Urry (1992) and Lister \& Marscher (1997), \[\Phi (L,\;z)
= \Phi (L/\exp [T(z)/\tau ],\;z = 0),\] where $\tau$ is the evolution parameter in unit
of Hubble time.

Using the equations of the beamed luminosity function (see details
in Appendix B), we can estimate the parameters in the intrinsic
luminosity function, the Lorentz factor distribution and the
evolution form by the maximum likelihood method. We combine the
data from ``VLBA 2 cm survey" and ``MOJAVE survey" (Figure 3). We
again divide the sample into low and high redshift groups by
$z=0.1$. The values of the best estimated parameters from high
redshift group are  $a = - 1.73_{ - 0.28}^{ + 0.28}$, $ A = 2.48_{
- 0.09}^{ + 0.09} $,  $ \tau  = 0.29_{ - 0.01}^{ + 0.03} $
 and $
\log (L_1 ({\rm{W/Hz}})) = 26.25_{ - 0.12}^{ + 0.03} $; the
results from the low redshift group are $a =  - 1.5_{ - 0.6}^{ +
0.6}$, $A = 2.8_{ - 1.0}^{ + 1.0}$
 and $
\log (L_1 ({\rm{W/Hz}})) = 24.39_{ - 0.40}^{ + 0.09} $ (no
evolution is assumed for the low redshift group, and $1\sigma$
error for each parameter of concern is shown). The value of $\log
(L_2 ({\rm{W/Hz}}))$ is fixed at 34, since higher values barely
change the results. Figure 4 shows the confidence region for $(a,
A)$ by fixing other parameters at the best estimated values.
Although the confidence regions in Figure 4 for the low and high
redshift groups are consistent within the 68\% confidence region,
the lower intrinsic luminosity limits of the two groups are
different significantly. As also shown in Figure 4, the result
from the luminosity data for the low redshift group is only
marginally consistent with the result from the apparent velocity
data within the 68\% confidence region. This indicates that our
current model do not describe the low redshift group very
appropriately. Actually, if we adopt the whole sample, the result
of the Lorentz factor distribution is unreasonable, i.e. $a>0$ is
required to obtain the maximum value of the estimator, which is
ruled out by previous studies. For example, if we adopt $ \log
(L_1 ({\rm{W/Hz}})) = 24.39$ indicated by the low redshift group,
the values of best estimated parameters of the whole sample are
$a=30.6$, $A=3.5$ and $\tau=1.0$. The K-S test is performed to
estimate the goodness of fit; the corresponding probabilities of
the apparent velocity, the observed luminosity and the redshift
distribution are $10^{-22}$, 10\% and 7\%, respectively.
Therefore, the consistency is quite poor, especially for the
apparent velocity.
 If the lower intrinsic luminosity limit is adjusted as a free parameter,
  the value of $a$ will become larger and even without an upper limit, which is more
  inconsistent with the result from the apparent velocity data. Therefore, we find that
  the results from both the apparent velocity and the observed luminosity data indicate the low
  and high redshift groups are likely to be from different parent populations. Due to the limited
   size of the low redshift group, we only utilize the high redshift groups in further analysis.

 As shown in Figure 5, there is slight difference between the result obtained from the apparent
 velocity data and the observed luminosity function (within 68\% confidence region).
 This is mainly due to the simple evolution form used here. When performing the K-S test
 to estimate the goodness of fit, we find the result of the best estimated value of parameters
 could only be accepted marginally. The corresponding probabilities of the apparent velocity,
  the observed luminosity and the redshift distribution are 45\%, 56\% and 6\%, respectively. However,
   if we adjust the value of $\tau$ to the $1\sigma$ boundary ($\tau=0.32$), the result of K-S test could be improved.
   The corresponding probabilities of the apparent velocity, the observed luminosity and
   the redshift distribution are 45\%, 61\% and 20\%, respectively. The effect of a higher
   flux limit is only mild (see the inset in Figure 4).

\section{ Implication for the unification scheme of radio loud AGNs}
There are two key parameters in the unification scheme of radio loud AGNs. One is the
orientation of the sources, and another is the intrinsic luminosity of the sources. The
high luminosity population contains quasars and luminous radio galaxies (FR II), while
the low luminosity population contains BL Lac objects and less luminous radio galaxies
(FR I). On the other hand, the viewing angles of quasars and BL Lac objects are smaller
than that of radio galaxies.  Therefore, the superluminal motion is more common in the
aligned objects, and the observed luminosity is strongly affected by special
relativistic effects. However, the relativistic beaming effect of radio galaxies is
mild and the observed luminosity function of radio galaxies is supposed to be similar
to the intrinsic one. The discussion about the low luminosity population could be found
in Padovaani \& Urry (1990, 1991) and Urry et al. (1991), and the unification scheme
was reviewed in Urry \& Padovani (1995). Due to the limited sample size of low
luminosity population and unclear relationship between FR I and FR II, we focus our
discussion on the high luminosity population below.

Several papers investigated the luminosity function of AGNs. For
example, Urry \& Padovani (1995) found the evolution parameter
$\tau$
 of FR II and FSRQ are $0.26_{-0.10}^{+0.74}$ and $0.23_{-0.04}^{+0.07}$, respectively. Padovani \& Urry (1992) adopted a double
 power law to fit the observed luminosity function of FR II, which was identified as the
 parent luminosity function. The power law indices of low and high luminosity band were
 found to be
$B_1=2.48_{- 0.15}^{ + 0.15}$ and $B_2=3.9_{ - 0.7}^{ + 0.7}$, respectively. Here we
have determined simultaneously the intrinsic luminosity function and the
 Lorentz factor distribution of relativistic jets in AGNs by the maximum likelihood method.
 However, we find a single power law form of the intrinsic luminosity function
 is sufficient to describe the sample we used here. For comparison, we have
 re-analyzed the observed luminosity function of the FR II sample adopted in
  Padovani \& Urry (1992) by the maximum likelihood method (for completeness,
  only the sources with $z<1.3$ are used). To compare with our result, we also
  adopt a single power law ($\Phi (L_o ,z = 0) \propto L_o ^{ - A} $) and the pure luminosity evolution.
  The results
 are $A = 2.51_{ - 0.15}^{ + 0.15} $ and $\tau  = 0.40_{ - 0.09}^{ + 0.09}$. The corresponding probabilities
obtained by K-S test of the observed
   luminosity and the redshift distribution are 75\% and 80\%, respectively. Therefore,
   we find this simple model could describe the data well, and the result is consistent
   with the intrinsic luminosity function we have obtained in \S2. In previous studies (Padovani \& Urry
    1992; Lister \& Marscher 1997),
   it was assumed that the shape
of the observed luminosity function of FR II is the same with the intrinsic luminosity
function of the radio loud quasars. Therefore, our results prove the validity of this
assumption. The single
   power law form of the luminosity function may be somewhat simplified, and a more
   complex form will be investigated when a larger and more complete sample is available.
\section{ Discussions and conclusions}

We have determined simultaneously the intrinsic luminosity
function and the Lorentz factor distribution of relativistic jets
in AGNs by the maximum likelihood method, and have confirmed the
previous assumption about the shape of the intrinsic luminosity
fucntion. The result of the Lorentz factor distribution is also
consistent with the previous result. For example, Lister (1997)
claimed the index of the Lorentz factor distribution $a$ is
roughly between -1.75 and -1.5. However, he fixed the parameters
of the intrinsic luminosity as \sl a prior \rm
 condition.

In the previous studies, it has been assumed the Lorentz factor distribution is the
same at all redshift. However, we find the Lorentz factor distribution is much steeper
at low redshift $(z<0.1)$ with the result from apparent velocity data, though the
uncertainties of the results are large. This indicates that the low and high redshift
groups are likely to be from different parent populations, i.e. the dual-population
scheme (Jackson \& Wall 1999). The majority of low reshift sources are low luminosity
ones. They are not as energetic as the high luminosity quasars. Therefore, the most
extremely relativistic jets are relatively rare in this population. However, this is
not indicated by the results from the observed luminosity function. As discussed in
\S2, this is likely to be due to the relatively simple model applied here. Further more
detailed study could be performed when the sample is large and complete enough.

We assume the Lorentz factor is independent of the intrinsic
luminosity, i.e. the $L-\gamma$-independent (LGI) model. Lister \&
Marscher (1997) investigated a particular $L-\gamma$-dependent
(LGD) scenario ($L \propto \gamma ^\xi$). They found the
predictions of the best-fit LGD model were very similar to the
best-fit LGI model but predicted very few high viewing angle
sources compared with the Caltech-Jodrell Bank sample. We
speculate this may be due to the somewhat arbitrary form of the
relation between the Lorentz factor and the intrinsic luminosity.
Actually, even when the LGI model is employed, there should be
some correlation of the Lorentz factor and the intrinsic
luminosity in the observed sources due to the higher luminosity
threshold at higher redshift.

Besides the shape of the intrinsic luminosity function, the space density of different
types of AGNs should match with each other according to the unified scheme. However,
both of the space densities of FR II and FSRQ evolve with redshift significantly, and
the value of the dividing angle of different types of AGNs may also be related with the
luminosity and redshift (e.g. Willott et al. 2000; Arshakian 2005). It is therefore
quite complicated to demonstrate the consistency of space densities of different types
of AGNs within the context of the unification scheme, and thus beyond the scope of the
current work.

\acknowledgments{\bf Acknowledgement:} We thank ``MOJAVE survey" sharing their data on
the website. We also thank the anonymous referees for valuable suggestions and
comments, in particular the referee who suggested us to expand the previous submission
to ApJL and then to re-submitting it to the ApJ main journal. SNZ acknowledges partial
funding support by the Ministry of Education of China, Directional Research Project of
the Chinese Academy of Sciences under project no. KJCX2-YW-T03, and by the National
Natural Science Foundation of China under project no. 10521001.

\appendix

\section{Calculation about the apparent velocity distribution}

 \label{appendix}
The approach presented here is similar to that in Vermeulen \& Cohen (1994). The
apparent velocity probability density function (pdf) $p(\beta_{app})$ could be obtained
by the differentiation of the cumulative distribution function
$P(\beta_{app}^{'}\leq\beta_{app})$,

\begin{equation}
p(\beta _{app} ) = \frac{{dP(\beta _{app}^{'}  \le \beta _{app}
)}}{{d\beta _{app} }}\;,
\end{equation}
and
\begin{equation}
P(\beta _{app}^{'}  \le \beta _{app} ) = 1 - P(\beta _{app}^{'}
>\beta _{app} ) = 1 - \int_{\gamma _l }^{\gamma _u } {d\gamma
\int_{\theta _l }^{\theta _u } {p(\theta ,\gamma )d\theta } }\;,
\end{equation}
where \[\begin{array}{l}
 \gamma _l  = \sqrt {\beta _{app}^2  + 1}\;,  \\
 \gamma _u  = \gamma _{\max }\;,  \\
 \end{array}\]

 \[\begin{array}{l}
 \cos \theta _l  = \frac{{\beta _{app}^2 \gamma  + \sqrt {\gamma ^2  - 1 - \beta _{app}^2 } }}{{(1 + \beta _{app}^2 )\sqrt {\gamma ^2  - 1} }} \;,\\
 \cos \theta _u  = \left\{ {\begin{array}{*{20}c}
   {0\;,\;\;\;\;\;\;\;\;\;\;\;\;\;\;\;\;\;\;\;\;\;\;\;\;\beta _{app}  \le \beta }\;,  \\
   {\frac{{\beta _{app}^2 \gamma  - \sqrt {\gamma ^2  - 1 - \beta _{app}^2 } }}{{(1 + \beta _{app}^2 )\sqrt {\gamma ^2  - 1} }}\;,\;\;\;\beta _{app}  \ge \beta }\;,  \\
\end{array}} \right. \\
 \end{array}\]
and $\gamma=1/\sqrt{1-\beta^2}$ .

Due to the Doppler boosting effect, $p(\theta ,\gamma )$ depends on the observed
luminosity $L_o=L\delta^n$. We assume pdf of $L$ has a power law form, \[p(L) =
\frac{{qL^{ - (q+1)} }}{{(L_{\min }^{ - q} - L_{\max }^{ - q} )}}\;,\] then
\[
\begin{array}{l}
 p(L_o |\;\theta ,\gamma )dL_o  = p(L = [\gamma (1 - \beta \cos \theta )]^n L_o )dL \\
 \;\;\;\;\;\;\;\;\;\;\;\;\;\;\;\; \;\;\;\;\;\;= \frac{{q\{ [\gamma (1 - \beta \cos \theta )]^n L_o \} ^{ - (q + 1)} }}{{(L_{\min }^{ - q}  - L_{\max }^{ - q} )}}[\gamma (1 - \beta \cos \theta )]^n dL_o  \\
 \;\;\;\;\;\;\;\;\;\;\;\;\;\;\;\;\;\;\;\;\;\; = \frac{{q[\gamma (1 - \beta \cos \theta )]^{ - nq} }}{{L_o ^{q + 1} (L_{\min }^{ - q}  - L_{\max }^{ - q} )}}dL_o \;. \\
 \end{array}
\]
Therefore,
\begin{equation}
\begin{array}{l}
 p(\theta ,\gamma |\;L_o ) = p(L_o |\;\theta ,\gamma )p(\theta )p(\gamma )/[\int_1^{\gamma _{\max } } {p(\gamma )d\gamma \int_0^{\pi /2} {p(\theta )} } p(L_o |\theta ,\gamma )d\theta ] \\
 \;\;\;\;\;\;\;\;\;\;\;\;\;\;\;\; = [\gamma (1 - \beta \cos \theta )]^{ - nq} \sin \theta p(\gamma )/[\int_1^{\gamma _{\max } } {p(\gamma )\frac{1}{{\gamma \beta }}\frac{{[\gamma (1 - \beta )]^{ - nq + 1}  - \gamma ^{ - nq + 1} }}{{nq - 1}}d\gamma ]}\;.  \\
 \end{array}
\end{equation}
Note that there is not $L_o$ in equation (A3). Therefore,
\[p(\theta ,\gamma |\;L_o )=p(\theta ,\gamma)\;.\]
Substitute equation (A3) into equation (A2), we have
\[
\begin{array}{l}
 P(\beta _{app}^{'}  > \beta _{app} ) = \int_{\gamma _l }^{\gamma _u } {d\gamma \int_{\theta _l }^{\theta _u } {p(\theta ,\gamma )d\theta } }  \\
 \;\;\;\;\;\;\;\;\;\;\;\;\;\;\;\;\;\;\;\;\;\;\; = C\int_{\gamma _l }^{\gamma _u } {p(\gamma )d\gamma \int_{\cos \theta _u }^{\cos \theta _l } {[\gamma (1 - \beta u)]^{ - nq} du} }  \\
 \;\;\;\;\;\;\;\;\;\;\;\;\;\;\;\;\;\;\;\;\;\;\; = C\int_{\gamma _l }^{\gamma _u } {\frac{{p(\gamma )}}{{\gamma \beta }}d\gamma \frac{{[\gamma (1 - \beta \cos \theta _l )]^{ - nq + 1}  - [\gamma (1 - \beta \cos \theta _u )]^{ - nq + 1} }}{{nq - 1}}}  \\
 \;\;\;\;\;\;\;\;\;\;\;\;\;\;\;\;\;\;\;\;\;\;\; = C\int_{\gamma _l }^{\gamma _u } {\{ \frac{{p(\gamma )}}{{\gamma \beta }}[\gamma (1 - \beta \cos \theta _l )]^{ - nq + 1}  - [\gamma (1 - \beta \cos \theta _u )]^{ - nq + 1} \} } d\gamma  \;.\\
 \end{array}
\]
To be consistent with the notation in Appendix B, we denote
$A=q+1$.

 Using equation (A1), we could obtain the pdf of the apparent
velocity distribution. In this paper we assume $p(\gamma)$ is a power law, i.e.
$p(\gamma)=C \gamma^a$. The examples
 of different values of $a$ are shown in Figure 6

\section{Calculation about the apparent luminosity function}
The method presented here is the same with that in Lister (2003). Here we still list
the main equations for clarity.

To calculate the apparent luminosity function, we should know pdf of $\delta$. We
assume the orientation of jets is random, i.e. $p(\theta)\propto\sin(\theta)$.
Therefore, we have
\[
p(\delta ) = \delta ^{ - 2} \int_{c(\delta )}^{\gamma _{\max } }
{\frac{{p(\gamma )}}{{\sqrt {\gamma ^2  - 1} }}} d\gamma \;,\]
where \[c(\delta ) = \left\{ {\begin{array}{*{20}c}
   {\delta ^{ - 1}, \;\;\;\;\;\;\;\;\;\;\;\;\;\;\;\;\;\;\;\gamma _{\max }^{ - 1}  \le \delta  \le 1}\;,  \\
   {\frac{{1 + \delta ^2 }}{{2\delta }}\;\;,1 \le \delta  \le \delta _{\max }  = \gamma _{\max } (\beta _{\max }  + 1)}\;,  \\
\end{array}} \right.\]

We assume the form of the intrinsic luminosity function is
\[ \phi(L) = kL^{ - A}  \;  (L_{1}\leq L \leq L_{2})\;. \]
Since \[ \Phi (L_{\rm{o}} ,0) = \int {p(L_o ,\delta )d\delta }  =
\int {p(\delta )\phi (L)\frac{{dL}}{{dL_o }}d\delta } \;,\] and
$L_o=L\delta^n$, we have
\[
\Phi (L_{\rm{o}}) = kL_{\rm{o}} ^{ - A} \int_{a(L_{\rm{o}}
)}^{b(L_{\rm{o}} )} {\delta ^{ - n + nA - 2} d\delta
\int_{c(\delta )}^{\gamma _{\max } } {\frac{{p(\gamma )}}{{\sqrt
{\gamma ^2  - 1} }}} d\gamma } \;,\]

where \[\begin{array}{l}
 \;a(L_o ) = \gamma _{\max }^{ - 1}, \;\;\;\;b(L_o ) = (L_o /L_1 )^{1/n} \;\;(L_1 \gamma _{\max }^{ - n}  \le L_o  < L_1 \delta _{\max }^n )\;,\;\;\;\;\;\;\\
 \;a(L_o ) = \gamma _{\max }^{ - 1}, \;\;\;\;b(L_o ) = \delta _{\max } \;\;\;\;\;\;\;\;\;\;\;(L_1 \delta _{\max }^n  \le L_o  < L_2 \gamma _{\max }^{ - n} )\;,\;\\
 \;a(L_o ) = (L_o /L_2 )^{1/n}, \;\;\;\;b(L_o ) = \delta _{\max } \;\;(L_2 \gamma _{\max }^{ - n}  \le L_o  \le L_2 \delta _{\max }^n ) \;,\\
 \end{array}\]

In this paper we assume $p(\gamma)$ is a power law, i.e.
$p(\gamma)=C \gamma^a$. The
 results of different values of $a$ are shown in Figure 7. This figure
 is similar to the Figure 3 in Lister (2003) (We mention in passing that there are
 some typos in the figure caption.).

%% The reference list follows the main body and any appendices.
%% Use LaTeX's thebibliography environment to mark up your reference list.
%% Note \begin{thebibliography} is followed by an empty set of
%% curly braces.  If you forget this, LaTeX will generate the error
%% "Perhaps a missing \item?".
%%
%% thebibliography produces citations in the text using \bibitem-\cite
%% cross-referencing. Each reference is preceded by a
%% \bibitem command that defines in curly braces the KEY that corresponds
%% to the KEY in the \cite commands (see the first section above).
%% Make sure that you provide a unique KEY for every \bibitem or else the
%% paper will not LaTeX. The square brackets should contain
%% the citation text that LaTeX will insert in
%% place of the \cite commands.

%% We have used macros to produce journal name abbreviations.
%% AASTeX provides a number of these for the more frequently-cited journals.
%% See the Author Guide for a list of them.

%% Note that the style of the \bibitem labels (in []) is slightly
%% different from previous examples.  The natbib system solves a host
%% of citation expression problems, but it is necessary to clearly
%% delimit the year from the author name used in the citation.
%% See the natbib documentation for more details and options.

\clearpage

\begin{figure}
\begin{center}
\epsscale{.3}
\includegraphics[width=13cm]{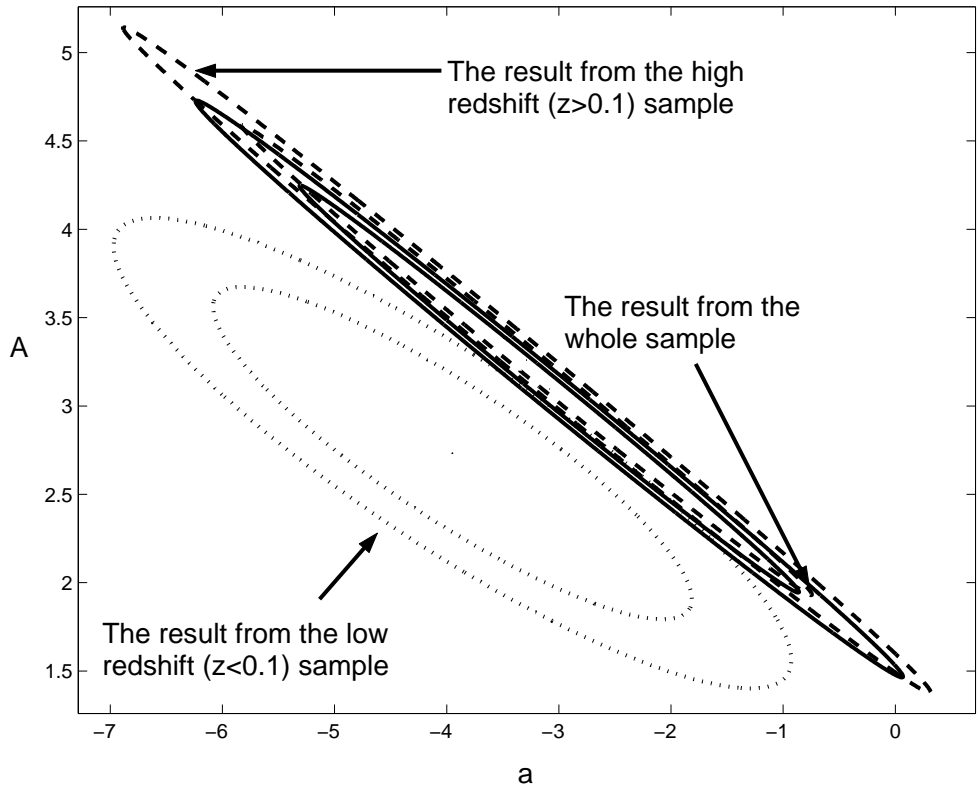}
\end{center}
\caption{Inferred parameters from the apparent velocity data. The
confidence regions are results obtained with the whole sample
(solid lines), $z>0.1$ sources (dashed lines), and $z<0.1$ sources
(dot lines), respectively. The 68\% and 90\% confidence regions
are shown for each case.}
\end{figure}

\clearpage

\begin{figure}
\begin{center}
\includegraphics[width=13cm]{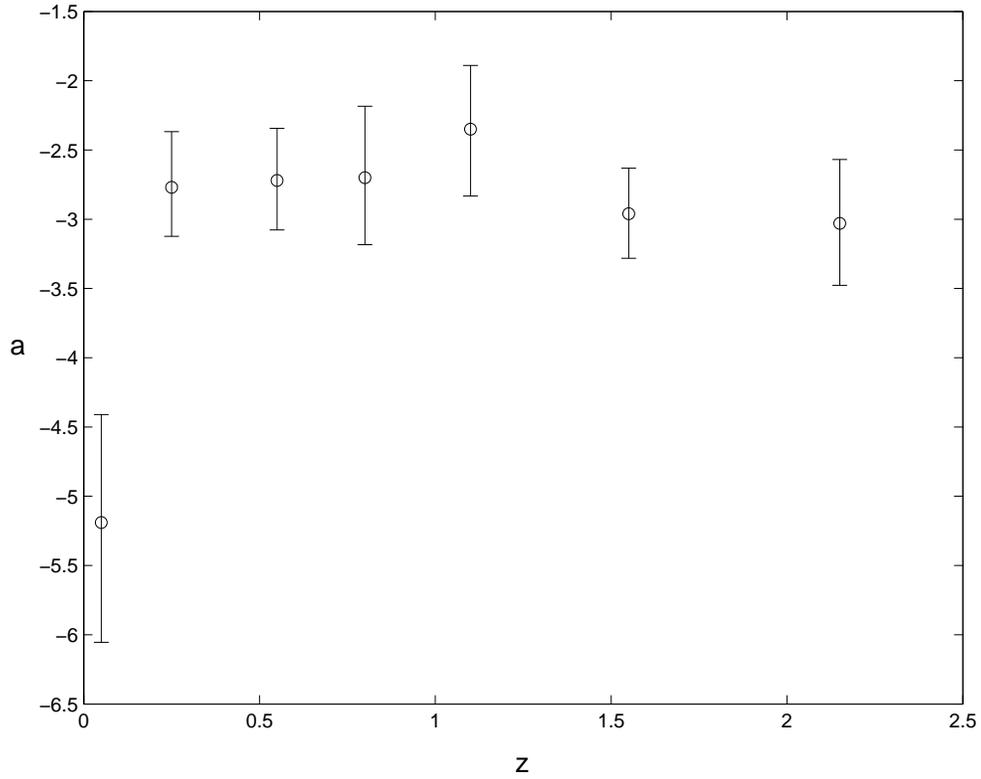}
\end{center}
\caption{Evolution of the index of the Lorentz factor distribution by fixing $A=3.10$.
From low to high reshift, there are 13, 17, 17, 13, 17, 17 and 10 sources in each
redshift bin, respectively. The $1\sigma$ errors are shown for each redshfit bin.}
\end{figure}

\clearpage

\begin{figure}
\begin{center}
\includegraphics[width=13cm]{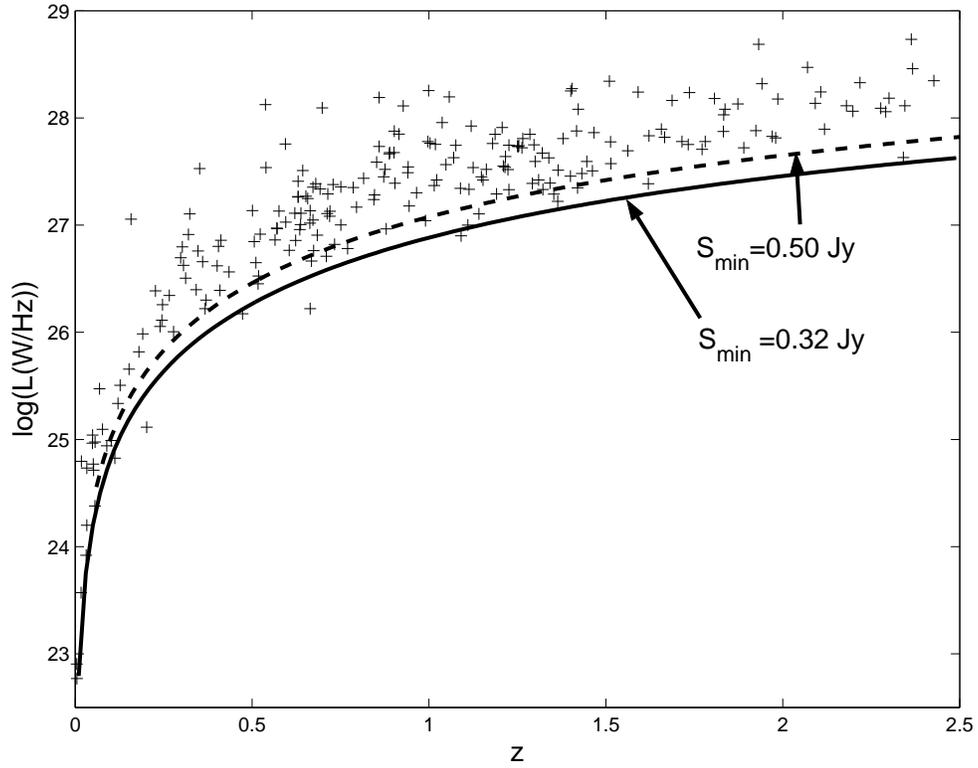}
\end{center}
\caption{The luminosity and redshift sample (MOJAVE survey). The
solid and dashed lines are the flux limit corresponding to 0.32 Jy
and 0.50 Jy, respectively. For completeness, only the sources with
$z<2.5$ are included.}
\end{figure}

\clearpage

\begin{figure}
\begin{center}
\includegraphics[width=13cm]{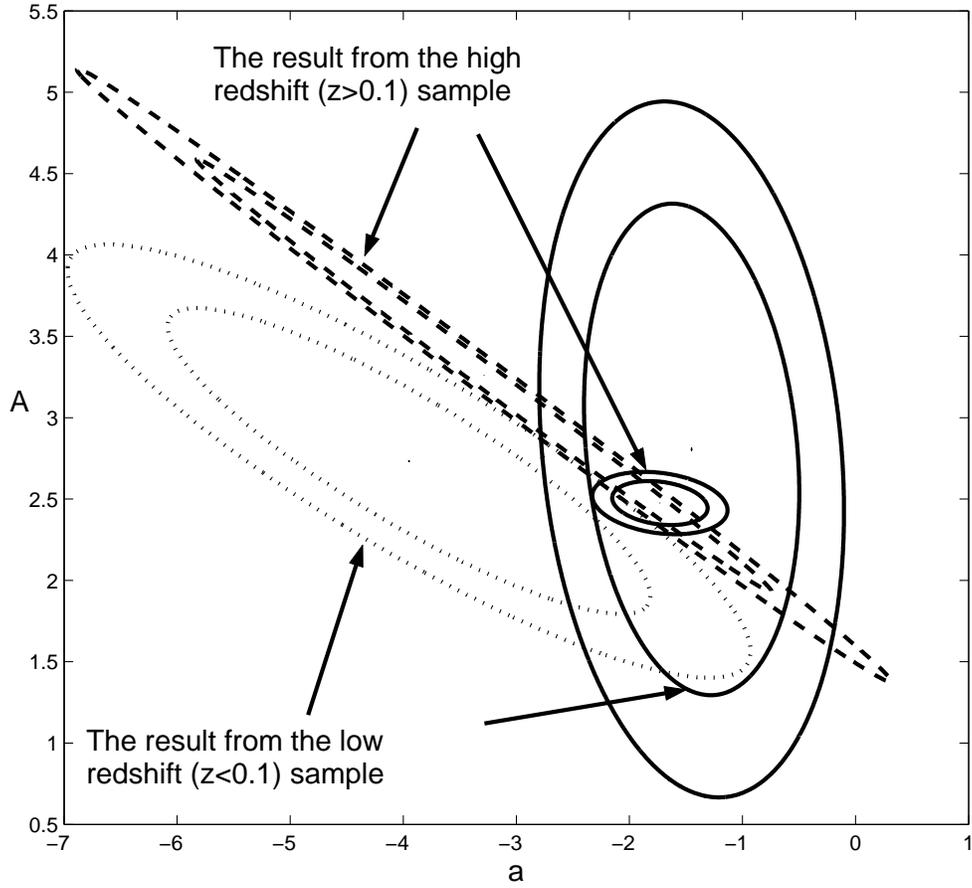}
\end{center}
\caption{The results from the observed luminosity and redshift
(solid lines). For comparison, the results from the apparent
velocity are also shown (dashed lines). The 68\% and 90\%
confidence regions are shown for each case.}
\end{figure}

\clearpage

\begin{figure}
\begin{center}
\includegraphics[width=13cm]{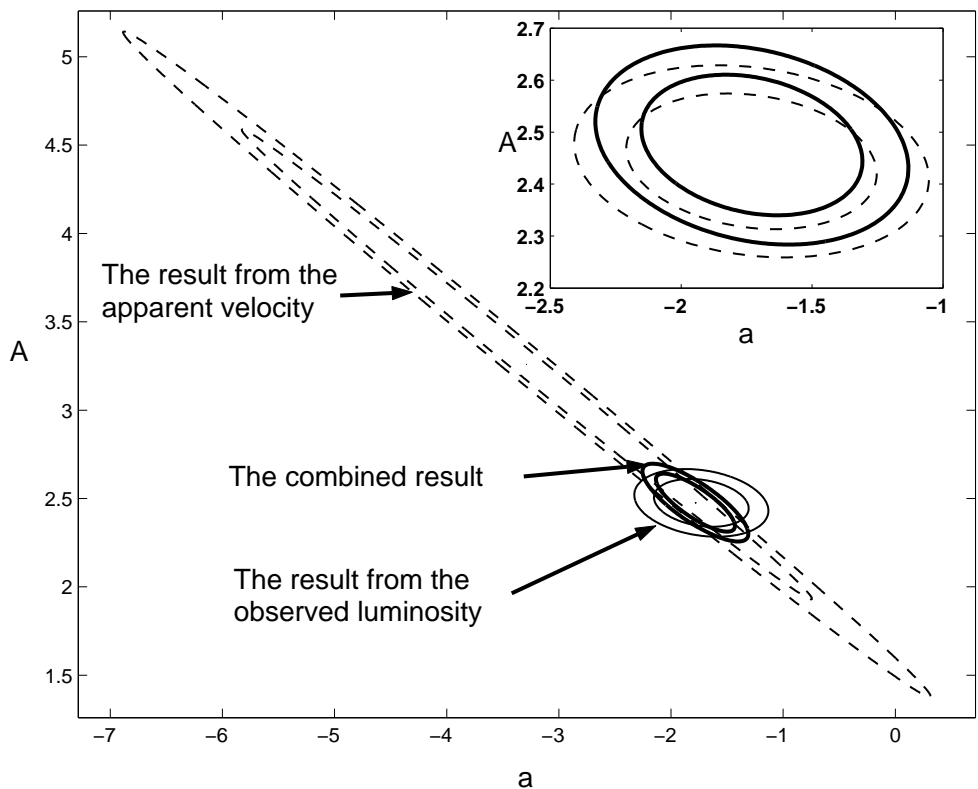}
\end{center}
\caption{The thin solid lines are the results from the observed luminosity
 and the redshift distribution. The dashed lines are the results from the
 apparent velocity distribution ($z>0.1$). The thick solid lines are the combined
  results. The inset shows the results from the observed luminosity and the redshift
   distribution for $S_{min}=0.32$ (solid lines) and $S_{min}=0.50$ (dashed lines). The 68\% and 90\% confidence regions are shown for each case.}
\end{figure}

\clearpage

\begin{figure}
\begin{center}
\includegraphics[width=13cm]{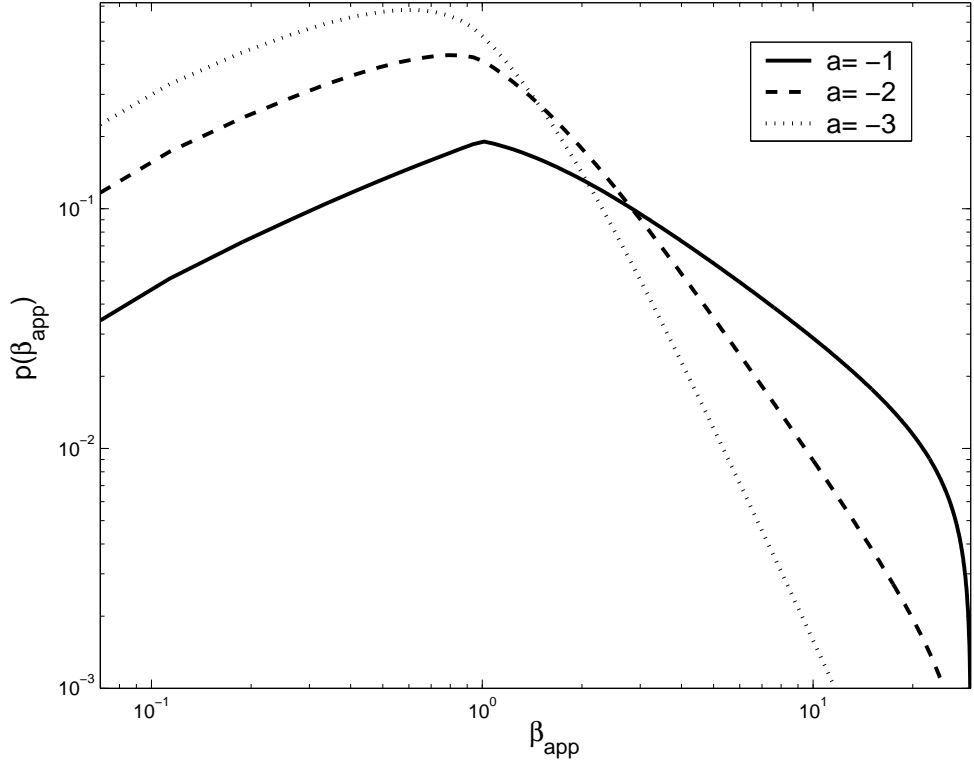}
\end{center}
\caption{The pdf of the apparent velocity distribution with
$p(\gamma)\propto\gamma^a (1<\gamma<30)$ and $n=2$. Three cases
are shown, i.e. $a=-1$ (solid line), $a=-2$ (dashed line) and
$a=-3$ (dot line).}
\end{figure}

\clearpage

\begin{figure}
\begin{center}
\includegraphics[width=13cm]{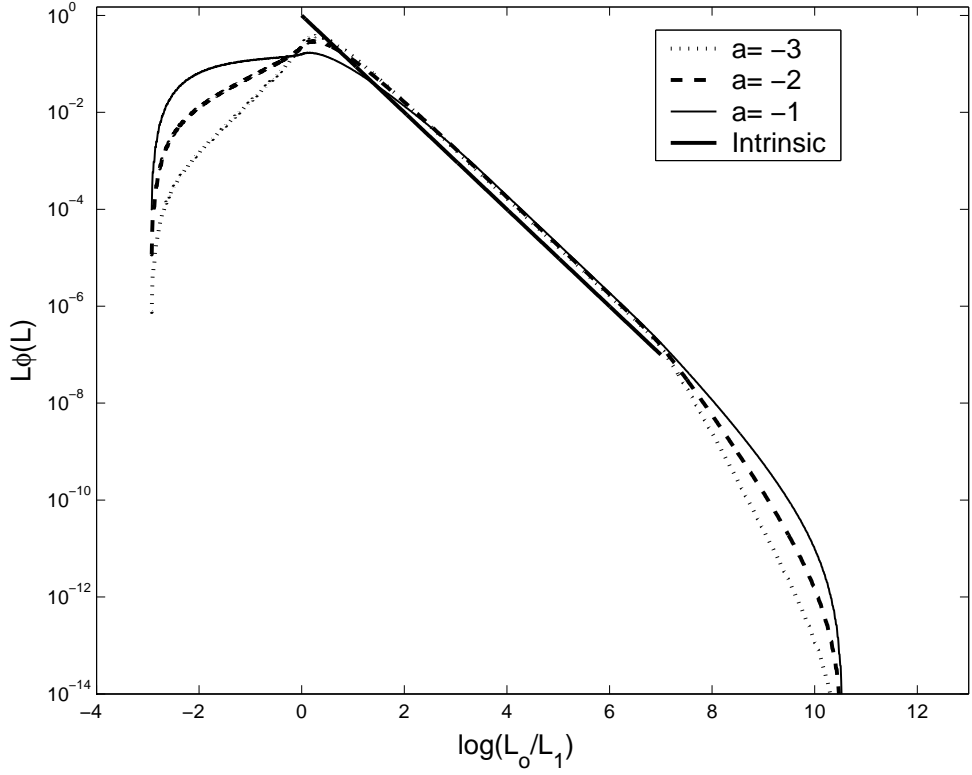}
\end{center}
\caption{The observed luminosity function with
$p(\gamma)\propto\gamma^a (1<\gamma<30)$ and $n=2$. The thick
solid line is the intrinsic luminosity function, i.e.
$\phi(L)\propto\L^{-2} \;(L_1<L<10^7 L_1)$. Three cases are shown,
i.e. $a=-1$ (thin solid line), $a=-2$ (dashed line) and $a=-3$
(dot line).}
\end{figure}
\end{document}